\documentclass[useAMS,usenatbib,onecolumn]{mn2e}

\usepackage{graphicx}
\usepackage{url}
\usepackage{epsfig}
\usepackage{color,amsmath,amssymb}

\renewcommand{\arraystretch}{0.8}

\newcommand{\3}{$_{3}$}
\newcommand{\4}{$_{4}$}
\newcommand{\cm}{cm$^{-1}$}

\newcommand{\Cvvv}{${\mathcal C}_{3{\rm v}}$}

\newcommand{\schr}{Schr\"{o}dinger}

\title{ExoMol line lists VII: The rotation-vibration spectrum of phosphine
up to 1500~K}

\date{\today}
\author[Sousa-Silva, Al-Refaie, Tennyson and Yurchenko]{\large
Clara Sousa-Silva, Ahmed F. Al-Refaie, Jonathan Tennyson and Sergei N. Yurchenko
\\
Department of Physics and Astronomy, University College London, London WC1E 6BT,
UK}
\date{Accepted XXXX. Received XXXX; in original form XXXX}

\pagerange{\pageref{firstpage}--\pageref{lastpage}} \pubyear{2014}

\begin{document}

\maketitle

\begin{abstract}

  A comprehensive hot line list is calculated for $^{31}$PH$_3$ in its
  ground electronic state. This line list, called SAlTY, contains
  almost 16.8 billion transitions between 7.5 million energy levels
  and it is suitable for simulating spectra up to temperatures of
  1500~K. It covers wavelengths longer than 1~$\mu$m and includes all
  transitions to upper states with energies below $hc \cdot
  18\,000$~cm$^{-1}$ and rotational excitation up to $J=46$. The line
  list is computed by variational solution of the Schr\"{o}dinger
  equation for the rotation-vibration motion employing the
  nuclear-motion program TROVE. A previously reported {\it ab initio}
  dipole moment surface is used as well as an updated `spectroscopic'
  potential energy surface (PES), obtained by refining an existing
  \textit{ab initio} surface through least-squares fitting to the
  experimentally derived energies. Detailed comparisons with other
  available sources of phosphine transitions confirms SAlTY's accuracy
  and illustrates the incompleteness of previous experimental and
  theoretical compilations for temperatures above 300 K. Atmospheric
  models are expected to severely underestimate the abundance of
  phosphine in disequilibrium environments, and it is predicted that
  phosphine will be detectable in the upper troposphere of many
  substellar objects. This list is suitable for modelling atmospheres
  of many astrophysical environments, namely carbon stars, Y dwarfs, T
  dwarfs, hot Jupiters and solar system gas giant planets. It is
  available in full as supplementary data to the article and at
  \url{www.exomol.com}.

\end{abstract}
\begin{keywords}
molecular data; opacity; astronomical data bases: miscellaneous; planets and
satellites: atmospheres; stars: low-mass; stars: brown dwarfs.
\end{keywords}

\label{firstpage}

\section{Introduction}

Over the past twenty years, discoveries of planets, stars and substellar objects have
demonstrated the enormous diversity of astrophysical bodies in the Universe. Thanks to
modern techniques and technology, it is now possible to study the atmospheres of these
objects and retrieve some knowledge of their composition, structure and dynamics.
However, these are extremely complex systems to model and accurate observations are still
difficult to perform. In particular, it is crucial for the correct characterisation of
these atmospheres that complete and accurate descriptions of the molecules that comprise them
are available. Here phosphine, or PH$_{3}$, is considered.

Phosphorus is the one of the most abundant chemically reactive volatile elements in a
solar type system (with S, after H, C, N and O). Although phosphorus has considerably
smaller cosmic abundances than H, O, C or N, it is predicted to have an important role in
atmospheric chemistry and dynamics. Phosphorus is not particularly common in the universe but it is ubiquitous and is
important for most essential biochemical functions. Due to its role as biogenic particle,
phosphorus could potentially be used in the search for extinct or extant life in other
planets \citep{05MaXXXX.PH3}.

 A large fraction of the existing phosphorus
in various astronomical environments is expected to be found in the form of
phosphine, or PH$_{3}$ \citep{14AgCeDe.PH3,06ViLoFe.PH3}. Phosphine is an extremely
toxic, semi-rigid, relatively stable, oblate, symmetric-top molecule, and has so far been
detected in the lower troposphere of the Earth, the atmospheres of Jupiter and Saturn, and in
the carbon star envelope IRC +10216 \citep{75PrLeXX.PH3,98EdAtTr.PH3,09FlOrTe.PH3,
14AgCeDe.PH3, 92TaLaLe.PH3}.

Partially due to the lack of strong absorption from CH$_4$, the  5~$\mu$m spectral window
is a region of low opacity in Jupiter, Saturn and the Earth's atmosphere. Alongside other
molecules (CO, GeH$_4$, AsH$_3$), PH$_3$ is partly responsible for the continuum opacity
in this region and has detectable absorption bands \citep{97NoMaXX.PH3}, specifically the
$2\nu_2$ overtone bending band. In Jupiter, phosphorus is found at approximately
tropospheric solar abundances (1.1~ppm) in the form of phosphine, and is over four
times more abundant in Saturn (4.5~ppm), also as phosphine \citep{97NoMaXX.PH3,
03OwEnXX.PH3, 09MoMaLe.PH3, 06ViLoFe.PH3}.  This agrees with models of core accretion
that expect Saturn to have a considerably larger fraction of ice to gas than Jupiter. In
both planets, this abundance of a disequilibrium species like phosphine at observable
levels of the atmosphere reflects both a large reservoir in deep regions of the planets
and the strength of the convective transport of the molecule.

 In local thermochemical equilibrium (LTE), P$_{4}$O$_{6}$
would be expected to dominate the upper atmospheres of planets with approximate solar abundances. However, almost no
P$_{4}$O$_{6}$ is found in these environments because phosphine quenching and rapid
vertical mixing from regions where PH$_3$ is dominant leads to high disequilibrium
abundances. This quenching occurs because timescales for the conversion of PH$_{3}$  to
P$_{4}$O$_{6}$ are much larger than the convective timescales, so the production of
P$_{4}$O$_{6}$ is kinetically inhibited \citep{14MoXXXX.exo, 06ViLoFe.PH3,
96FeLoXX.dwarfs}. Disequilibrium chemistry is
expected to lead to phosphine being orders of magnitude more abundant than predicted by
equilibrium models, both for environments where T $\geq2000 K$ and at the uppermost
atmosphere of most substellar objects \citep{14MoXXXX.exo}.

The thermal decomposition of phosphine at high temperatures has been
studied for a long time, starting with \citet{24HiToXX}. Recently
\citet{06ViLoFe.PH3} discussed at length the pressure and temperature
parameters at which phosphine is expected to be found  in solar
composition and LTE environments. At 1 bar, phosphine is predicted to
be a dominant phosphorus carrying molecule in environments with
temperatures up to about 1500 K, and it is expected to contribute
to the composition of atmospheres up to about 2500 K if the
pressure is above 3 bar.

 The detection of phosphine can therefore be used as a chemical probe for deep
layers of both Jupiter and Saturn's atmosphere, as it is only present at the top layers
through vertical mixing. Analysis of future observations of Jupiter from NASA's TEXES,
NASA's Juno and ESA's JUICE will require accurate data on phosphine, and past
observations of Saturn from Cassini/VIMS and Cassini/CIRS will gain from further
understanding its spectrum \citep{14EnGrDr.PH3, 09FlOrTe.PH3}. Suspected inaccuracies
found in regions of the existing phosphine data may be behind misinterpretation of
previous astronomical data \citep{14DeKlSa.PH3}. Additionally, JUNO will cover the full
latitude and longitude of Jupiter and its analysis of the polar atmosphere of Jupiter
should confirm a depletion of phosphine in the polar vortices analogous to Saturn's
\citep{09FlOrTe.PH3}. Titan has also been predicted to contain phosphine, but spectrum
recorded with Cassini CIRS failed to detect any, putting an upper limit of 1 ppb on the
phosphine abundance of its atmosphere \citep{13NiTeIr.PH3}. Similar unsuccessful
detections of phosphine in Neptune and Uranus suggest that the abundance of phosphorus
on these planets is sub solar \citep{09MoMaLe.PH3}, with an upper limit of 0.1 times
solar P/H abundance.

The existence of phosphine outside the Solar System has recently been confirmed by \citet{14AgCeDe.PH3} in the
carbon star envelope IRC +10216, with an abundance of $10^{-8}$ relative to H$_{2}$. This was done using the HIFI instrument on board Herschel. Together with HCP, PH$_{3}$ is one of the major phosphorus carriers
in the inner circumstellar regions.


There have been multiple predictions of phosphine in many other astrophysical
environments, but its presence and formation scenarios remain poorly understood and none
has been found outside the solar system other than the aforementioned discovery of
circumstellar phosphine in IRC +10216. This is due to a variety of factors, including the
sparsity of fundamental data which this work is rectifying. Below are some of the
expectations for further phosphine detection.

 A particular region of interest is the 4.1-5.1~$\mu$m window, which is found to be an
interval of very low opacities in both Jupiter and Saturn. It is therefore expected that detection of PH$_3$ in this region could be used as a
marker for vertical convection and a tracer for tropospheric dynamics in the upper
atmospheres of other astronomical bodies, namely Hot Jupiters and Brown Dwarfs.

In Y dwarfs, with $T_{\rm eff}\approx500$~K, phosphine should exhibit a strong feature in
the mid-infrared at 4.3~$\mu$m \citep{14MoMaFo.PH3}, where it is predicted to be the
dominant source of opacity. T dwarfs, with $T_{\rm eff}\leq1300$~K, will mostly be
dominated by other molecules (e.g. H$_2$, NH$_3$, CH$_4$, CO, H$_2$O and CO$_2$) but
phosphine can still have a significant contribution to the shape of the spectrum. The
NIRSpec near-infrared spectrograph on JWST will be able to collect spectra in the
2.9-5~$\mu$m region with sufficient sensitivity to detect the presence or absence of
phosphine, particularly for slightly cooler Y dwarfs with $T_{\rm eff}\leq450$~K
\citep{14MoMaFo.PH3}.

Phosphine is predicted to be the dominant phosphorus molecule in cool T dwarfs like
Gliese 229B where it is expected to carry all the phosphorus in the atmosphere to the
top layers, with approximately 0.6 ppm. PH$_3$  should be detectable (for resolutions
higher than 1.2 cm$^{-1}$) in the 4.45-5.40 $\mu$m spectral window of Gliese 229B,
particularly the 4.3~$\mu$m feature \citep{97NoMaXX.PH3, 96FeLoXX.dwarfs}. Models
estimate that, for regions where $T\geq$1155 K, the major phosphorus carrying gas in
Gliese 229B is phosphine, with rapid vertical mixing quenching its destruction
\citep{96FeLoXX.dwarfs}. As the temperature rises, PH$_{3}$ is converted to
P$_{4}$O$_{6}$, but it should still be present at observable heights.

In hotter environments, like HD209458b and L dwarfs, where phosphorus equilibrium
chemistry is approached, the phosphine abundances decrease significantly to about 50 ppb
\citep{06ViLoFe.PH3}. It is still possible to detect phosphine in these atmospheres, but
higher resolution is required to distinguish its features from that of neighbouring CO
bands \citep{97NoMaXX.PH3}. However, phosphine is expected to be the dominant phosphorus
carrying species at the observable atmosphere of hot T dwarfs and cool L dwarfs, with
effective temperatures between 1000 and 1400~K \citep{06ViLoFe.PH3}.

There are still disequilibrium effects that are not completely understood. For example,
observations of Saturn's upper troposphere found that phosphine abundances decreased with
depth \citep{14DeKlSa.PH3}, contradicting the proposal that phosphine is
dredged up through vertical mixing from deeper layers of the atmosphere. This depletion
is not understood, as there is no known formation mechanism for  phosphine at the
upper levels of Saturn's atmosphere that could lead to the abundances observed. Suggested
explanations for this inconsistency are a poor understanding of the aerosol opacity in
the VIMS observations and intensity uncertainties in the molecular data.

Future detections of phosphine can only be achieved by accurately modelling these atmospheres. This
requires an extensive knowledge of the temperature dependent spectrum of phosphine, which
can only be established with a comprehensive line list.
	

There are multiple sources of phosphine data
\citep{13Muxxxx.PH3,07KsXXXX.PH3,06TeBaBu.PH3,06CaPuXX.PH3,06BuSaKl.PH3,05WaChCh.PH3,04UlBeKo.PH3,04SaArBo.PH3,02UlBeKo.PH3,02SuXXXX.PH3,02BrSaKl.PH3,01HeZhHu.PH3,00FuLoXX.PH3,97AiHaSp.PH3,92TaLaLe.PH3,81TaDaGo.PH3,81PiPoCo.PH3,81BeBuPo.PH3,80BaMaNa.PH3,
77HeGoXX.PH3,74ChOkXX.PH3,71DaNeWo.PH3,69HeGoXX.PH3,51LoStXX.PH3,00WaShZh.PH3,01HeZhHu.PH3,03YuCaJe.PH3,05YuThPa.PH3,06YuCaTh.PH3,08OvThYu.PH3,08OvThYu2.PH3,09NiHoTy.PH3},
both experimental and theoretical, which are further described in our previous paper on
room temperature phosphine \citep{jt556}. Additionally, a recent paper by \citet{14DeKlSa.PH3}
improved the experimental phosphine spectrum for the 4.08--5.13~$\mu$m region, which, as mentioned above, is of particular importance due to it
being an interval of low opacity in gas giants and brown dwarfs \citep{9NoMaXX.PH3}. Most
of these experimental sources have already been incorporated in the HITRAN database
\citep{jt557} and collectively contain less than 30~000 lines, and no wavelength coverage
shorter than 2.7~$\mu$m.

Despite these efforts, no complete line list for hot phosphine exists in the literature. The
line lists mentioned above only cover low values of rotational excitations, a small
number of bands, and some regions are missing altogether. Often, contamination of gas
samples affects the measured intensities \citep{14DeKlSa.PH3}, and this can be an effect
that is hard to accurately compensate for. It is not unusual for experimental data with
extremely accurate line positions to have intensity uncertainties well above 10\%.
Additionally, all the experimental datasets mentioned above are designed to be used only
at room temperature and below. Consequently, they are insufficiently complete to be
appropriate for use in the characterisation of atmospheres of any astronomical bodies at
higher temperatures.

A theoretical line list for phosphine was previously computed by us \citep{jt556},
henceforth SYT. However, although SYT contains many orders of magnitude more lines than
any previous phosphine line list (137 million transitions between 5.6 million energy
levels), it too was only designed to be accurate for temperatures below 300~K, making it
unsuitable for most astronomical studies. Since HITEMP \citep{jt480}, HITRAN's
high-temperature database, does not include phosphine, SAlTY is the first line list
suitable for modelling phosphine spectra in environments up to 1500~K. The completeness
and coverage of the SAlTY line list makes it particularly suitable for studies of non-LTE
environments.

The current work is performed as  part of the ExoMol project. This project aims
 to provide line lists of spectroscopic transitions for key molecular
 species which are likely to be important in the atmospheres of
 extrasolar planets and cool stars; its aims, scope and methodology are
 summarised in \citet{jt528}. The SAlTY line list builds on previously released
 ones for polyatomic hydrides such as water \citep{jt378}, ammonia \citep{jt500}, formaldehyde \citep{jt598}, and
 methane \citep{jt564,jt572}. Finally we note that energy levels generated for
 this work have been used as part of  a study
 to compute thermodynamic properties of phosphine and ammonia, including
 its partition function, for temperatures up to 6000~K \citep{jt571}.

\section{Overview of the SA\lowercase{l}TY line list}
\label{s:overview}

In the following PH$_3$ and phosphine will refer to the main isotopologue $^{31}$PH$_3$,
since $^{31}$P is the only non-synthetic, stable isotope of phosphorus.

SAlTY is a catalogue of transitions, each characterised by a
frequency, its lower and upper energy level, Einstein coefficient and
quantum numbers. Together these fully describe the spectrum of the
phosphine molecule within the frequency range 0 -- 10 000~cm$^{-1}$,
or wavelengths longer than 1~$\mu$m. It contains 16~803~703~395
transitions between 7~480~690 energy levels below 18~000 cm$^{-1}$,
with rotational quantum number $J$ values up to 46.  The highest
energy state considered is 18~000~cm$^{-1}$ above the zero-point
energy (5~213.9280~cm$^{-1}$) for phosphine, as the intensity of
transitions to higher energy levels is too weak to be important, even
at temperatures of 1500~K.  Consequently, to ensure that the line list
is complete within the frequency range 0 -- 10~000~cm$^{-1}$, the
highest lower energy state considered is 8~000~cm$^{-1}$. All
transitions are within the ground electronic state of phosphine, since
the first excited electronic state is above the dissociation limit of
the molecule.

The strongest SAlTY transition has an Einstein-$A$ coefficient of $89.1$ s$^{-1}$ while the weakest
lines go down to $10^{-48}$ s$^{-1}$. In non-LTE environments even extremely
weak lines can play an important role in the shape of a spectrum. This is the reason
behind including transitions with extremely weak absorptions in SAlTY.

The final line list is presented in the ExoMol format \citep{jt548},
with a transition file ordered in increasing transition frequency and
an energy file. The former contains a description of each transition
by its upper (final) and lower (initial) energy level reference
numbers ($f$ and $i$), as well as the electric dipole transition
probability represented as an Einstein coefficient $A_{\rm if}$ in
s$^{-1}$. Using this information, the line intensity of each
transition can be calculated for any given temperature. The latter
connects each index with the description of the corresponding energy
level. Each energy level is described by the quantum numbers
associated with the molecular group symmetry \citep{98BuJexx.method},
$C_{ 3v}$(M) for PH$_3$ and total angular momentum $J$.  The quantum
numbers for XY$_3$ molecules are quite complicated and have recently
been discussed in detail for ammonia by \citet{jt546}. Our PH$_3$
quantum numbers follow the same general principles but are somewhat
simpler since we neglect the possibility of a tunnelling mode.

Each energy level is described by the following quantum numbers :
\begin{equation}
\label{e:quanta}
  n_1, n_2, n_3, n_4, {L_3}, {L_4}, L, \Gamma_{\rm vib}, J, K, \Gamma_{\rm rot},
\Gamma_{\rm tot},
\end{equation}
where $L_3= |l_3|, L_4= |l_4|, L= |l|, K = |k| $. Here the vibrational
quantum numbers $n_{1}$ (symmetric stretch), $n_{2}$ (symmetric bend),
$n_{3}$ (asymmetric stretch) and $n_{4}$ (asymmetric bend) correspond
to excitations of, respectively, the $\nu_{1}$, $\nu_{2}$, $\nu_{3}$
and $\nu_{4}$ modes. The doubly degenerate modes $\nu_{3}$ and
$\nu_{4}$ require additional quantum numbers $L_{3} = |l_3|$ and
$L_{4}=|l_4|$ describing the projections of the corresponding angular
momenta (see \citet{98BuJexx.method}).  The vibrational quantum number
$L=|l|$, characterizes the coupling of $l_3$ and $l_4$. The projection
of the total vibrational angular momentum $L$ is included to reduce
ambiguity in the description of the energy levels, as it was in SYT
\citep{jt556}. $\Gamma_{\rm rot}$, $\Gamma_{\rm vib}$, and
$\Gamma_{\rm tot}$ are, respectively, the symmetry species of the
rotational, vibrational and total internal wave-functions in the
molecular symmetry group $C_{\rm 3v}$(M), spanning $A_1$, $A_2$ and
$E$, where $E$ is two-fold degenerate. $J$ is the total angular
momentum and $K=|k|, k= -J,\ldots ,J$ is the projection of the total
angular momentum on the molecule fixed axis $z$. These twelve quantum
numbers reduce ambiguity to the assignment of the energy levels.
However, only $J$ and $ \Gamma_{\rm vib}$ are rigorous; at higher
energies, energy states cannot necessarily be assigned unambiguous
quantum labels. Apart from the quantum numbers we also provide the
largest eigen-coefficient used to produce the theoretical assignment,
see, for example, \citep{jt564}.

Since PH$_3$ transitions obey the strict selection rules $A_{1} \leftrightarrow
A_{2}  \;\; , E \leftrightarrow E$, and $\Delta{J}=0,\pm1$, no higher $J$ values were
considered because $J=45$ is the highest value of $J$ for which there are eigenvalues
exist below 8 000~cm$^{-1}$, which is the highest lower energy threshold used in SAlTY.

%
%

Excerpts from the energy and transition  files are given in tables~\ref{t:Energy-file}
and ~\ref{t:Transit-file}, respectively, with seven additional columns giving TROVE quantum numbers which are described in the supplementary data and by \citet{jt556}. As per the ExoMol
convention, these are named SAlTY.transitions and SAlTY.states. The complete
line list is freely available and can be downloaded from the Strasbourg data
centre, CDS, via ftp://cdsarc.u-strasbg.fr/pub/cats/J/MNRAS/ or from
the ExoMol website, www.exomol.com. The website also offers the opportunity to download cross
sections \citep{jt542}.


\begin{table}
\caption{\label{t:Energy-file}  Extract from the SAlTY Energy file.}
\footnotesize \tabcolsep=5pt
\renewcommand{\arraystretch}{1.0}
\begin{tabular}{crrrrrrrrrrrrrrrrrrrrr}
    \hline
    \hline
  1     &        2        &  3  &  4  &  5  &  6  &  7  &  8  &  9  & 10  & 11
& 12  & 13  & 14  & 15 & 16 & 17 & 18 & 19 & 20 & 21 & 22   \\
    \hline
  $N$    &  $\tilde{E}$   &  $g_{\rm tot}$  &  $J$   &   $\Gamma_{\rm tot}$   &   $K$   &   $\Gamma_{\rm rot}$  &   $L$   &   $n_1$   &  $n_2$   &   $n_3$   & $n_4$  &  $L_3$  &  $L_4$  &  $\Gamma_{\rm vib}$ & $|C_i|^2$ & $s_1$   &  $s_2$   &   $s_3$   & $b_4$ & $b_5$   &  $b_6$\\
    \hline
4770	&	18091.558811	&	8	&	0	&	2	&	0	&	1	&	3	&	0	&	3	&	6	&	1	&	2	&	1	&	2	&	1.00	&	1	&	3	&	2	&	3	&	0	&	1	\\
4771	&	18108.514162	&	8	&	0	&	2	&	0	&	1	&	9	&	0	&	2	&	5	&	4	&	5	&	4	&	2	&	1.00	&	1	&	2	&	2	&	4	&	2	&	0	\\
4772	&	18117.126945	&	8	&	0	&	2	&	0	&	1	&	3	&	0	&	3	&	6	&	1	&	2	&	1	&	2	&	1.00	&	1	&	3	&	2	&	3	&	0	&	1	\\
4773	&	18141.236226	&	8	&	0	&	2	&	0	&	1	&	0	&	7	&	1	&	1	&	1	&	1	&	1	&	2	&	1.00	&	0	&	0	&	8	&	0	&	0	&	2	\\
4774	&	18144.653263	&	8	&	0	&	2	&	0	&	1	&	3	&	0	&	4	&	5	&	2	&	5	&	2	&	2	&	1.00	&	1	&	2	&	2	&	2	&	4	&	0	\\
4775	&	18180.965326	&	8	&	0	&	2	&	0	&	1	&	3	&	1	&	5	&	4	&	1	&	2	&	1	&	2	&	1.00	&	0	&	3	&	2	&	0	&	3	&	3	\\
4776	&	18218.349734	&	8	&	0	&	2	&	0	&	1	&	0	&	0	&	4	&	4	&	4	&	4	&	4	&	2	&	1.00	&	1	&	2	&	1	&	0	&	5	&	3	\\
4777	&	18287.993223	&	8	&	0	&	2	&	0	&	1	&	3	&	7	&	0	&	1	&	2	&	1	&	2	&	2	&	1.00	&	0	&	0	&	8	&	0	&	1	&	1	\\
4778	&	18329.569862	&	8	&	0	&	2	&	0	&	1	&	6	&	0	&	6	&	1	&	9	&	1	&	5	&	2	&	1.00	&	0	&	0	&	1	&	0	&	15	&	0	\\
4779	&	18393.321746	&	8	&	0	&	2	&	0	&	1	&	3	&	0	&	5	&	1	&	10	&	1	&	4	&	2	&	1.00	&	0	&	0	&	1	&	0	&	15	&	0	\\
4780	&	18453.246434	&	8	&	0	&	2	&	0	&	1	&	3	&	0	&	5	&	1	&	10	&	1	&	4	&	2	&	1.00	&	0	&	0	&	1	&	0	&	15	&	0	\\
4781	&	18506.447815	&	8	&	0	&	2	&	0	&	1	&	6	&	0	&	4	&	1	&	11	&	1	&	7	&	2	&	1.00	&	0	&	0	&	1	&	15	&	0	&	0	\\
4782	&	18516.395845	&	8	&	0	&	2	&	0	&	1	&	3	&	6	&	0	&	3	&	0	&	3	&	0	&	2	&	1.00	&	8	&	0	&	1	&	0	&	0	&	0	\\
4783	&	18548.610530	&	8	&	0	&	2	&	0	&	1	&	3	&	0	&	3	&	1	&	12	&	1	&	2	&	2	&	1.00	&	0	&	0	&	1	&	15	&	0	&	0	\\
4784	&	18649.304702	&	8	&	0	&	2	&	0	&	1	&	6	&	0	&	4	&	1	&	11	&	1	&	7	&	2	&	1.00	&	0	&	0	&	1	&	15	&	0	&	0	\\
4785	&	18725.969827	&	8	&	0	&	2	&	0	&	1	&	0	&	8	&	0	&	1	&	1	&	1	&	1	&	2	&	1.00	&	9	&	0	&	0	&	0	&	0	&	1	\\
4786	&	19353.229098	&	8	&	0	&	2	&	0	&	1	&	6	&	0	&	7	&	1	&	9	&	1	&	7	&	2	&	1.00	&	0	&	0	&	1	&	16	&	0	&	0	\\
4787	&	19493.996332	&	8	&	0	&	2	&	0	&	1	&	9	&	0	&	6	&	1	&	10	&	1	&	8	&	2	&	1.00	&	0	&	0	&	1	&	0	&	0	&	16	\\
4788	&	19745.979241	&	8	&	0	&	2	&	0	&	1	&	0	&	8	&	1	&	1	&	1	&	1	&	1	&	2	&	1.00	&	9	&	0	&	0	&	0	&	0	&	2	\\
4789	&	19894.271775	&	8	&	0	&	2	&	0	&	1	&	3	&	8	&	0	&	1	&	2	&	1	&	2	&	2	&	1.00	&	9	&	0	&	0	&	0	&	1	&	1	\\
4790	&	20126.886789	&	8	&	0	&	2	&	0	&	1	&	3	&	7	&	0	&	3	&	0	&	3	&	0	&	2	&	1.00	&	9	&	0	&	1	&	0	&	0	&	0	\\
4791	&	1118.304691	&	8	&	0	&	3	&	0	&	1	&	1	&	0	&	0	&	0	&	1	&	0	&	1	&	3	&	1.00	&	0	&	0	&	0	&	0	&	0	&	1	\\
4792	&	2108.150565	&	8	&	0	&	3	&	0	&	1	&	1	&	0	&	1	&	0	&	1	&	0	&	1	&	3	&	1.00	&	0	&	0	&	0	&	0	&	0	&	2	\\
4793	&	2234.920254	&	8	&	0	&	3	&	0	&	1	&	2	&	0	&	0	&	0	&	2	&	0	&	2	&	3	&	1.00	&	0	&	0	&	0	&	0	&	1	&	1	\\
4794	&	2326.870042	&	8	&	0	&	3	&	0	&	1	&	1	&	0	&	0	&	1	&	0	&	1	&	0	&	3	&	1.00	&	0	&	0	&	1	&	0	&	0	&	0	\\
4795	&	3085.609104	&	8	&	0	&	3	&	0	&	1	&	1	&	0	&	2	&	0	&	1	&	0	&	1	&	3	&	1.00	&	0	&	0	&	0	&	0	&	2	&	1	\\
4796	&	3222.494320	&	8	&	0	&	3	&	0	&	1	&	2	&	0	&	1	&	0	&	2	&	0	&	2	&	3	&	1.00	&	0	&	0	&	0	&	3	&	0	&	0	\\
4797	&	3311.958593	&	8	&	0	&	3	&	0	&	1	&	1	&	0	&	1	&	1	&	0	&	1	&	0	&	3	&	1.00	&	0	&	0	&	1	&	0	&	0	&	1	\\
4798	&	3333.494686	&	8	&	0	&	3	&	0	&	1	&	1	&	0	&	0	&	0	&	3	&	0	&	1	&	3	&	1.00	&	0	&	0	&	0	&	2	&	1	&	0	\\
4799	&	3424.626917	&	8	&	0	&	3	&	0	&	1	&	1	&	1	&	0	&	0	&	1	&	0	&	1	&	3	&	1.00	&	1	&	0	&	0	&	0	&	0	&	1	\\
4800	&	3435.624836	&	8	&	0	&	3	&	0	&	1	&	2	&	0	&	0	&	1	&	1	&	1	&	1	&	3	&	1.00	&	0	&	0	&	1	&	0	&	0	&	1	\\
4801	&	4050.520058	&	8	&	0	&	3	&	0	&	1	&	1	&	0	&	3	&	0	&	1	&	0	&	1	&	3	&	1.00	&	0	&	0	&	0	&	3	&	0	&	1	\\
4802	&	4196.913916	&	8	&	0	&	3	&	0	&	1	&	2	&	0	&	2	&	0	&	2	&	0	&	2	&	3	&	1.00	&	0	&	0	&	0	&	0	&	2	&	2	\\
4803	&	4283.755426	&	8	&	0	&	3	&	0	&	1	&	1	&	0	&	2	&	1	&	0	&	1	&	0	&	3	&	1.00	&	0	&	0	&	1	&	0	&	1	&	1	\\
4804	&	4319.935065	&	8	&	0	&	3	&	0	&	1	&	1	&	0	&	1	&	0	&	3	&	0	&	1	&	3	&	1.00	&	0	&	0	&	0	&	4	&	0	&	0	\\
4805	&	4408.422285	&	8	&	0	&	3	&	0	&	1	&	1	&	1	&	1	&	0	&	1	&	0	&	1	&	3	&	1.00	&	1	&	0	&	0	&	0	&	0	&	2	\\
4806	&	4418.431045	&	8	&	0	&	3	&	0	&	1	&	2	&	0	&	1	&	1	&	1	&	1	&	1	&	3	&	1.00	&	0	&	0	&	1	&	0	&	0	&	2	\\
4807	&	4438.012026	&	8	&	0	&	3	&	0	&	1	&	2	&	0	&	0	&	0	&	4	&	0	&	2	&	3	&	1.00	&	0	&	0	&	0	&	0	&	2	&	2	\\
4808	&	4461.860840	&	8	&	0	&	3	&	0	&	1	&	4	&	0	&	0	&	0	&	4	&	0	&	4	&	3	&	1.00	&	0	&	0	&	0	&	2	&	1	&	1	\\
4809	&	4517.161189	&	8	&	0	&	3	&	0	&	1	&	1	&	0	&	0	&	1	&	2	&	1	&	0	&	3	&	1.00	&	0	&	0	&	1	&	0	&	0	&	2	\\
4810	&	4534.991076	&	8	&	0	&	3	&	0	&	1	&	2	&	1	&	0	&	0	&	2	&	0	&	2	&	3	&	1.00	&	1	&	0	&	0	&	0	&	1	&	1	\\
4811	&	4545.384096	&	8	&	0	&	3	&	0	&	1	&	1	&	0	&	0	&	1	&	2	&	1	&	2	&	3	&	1.00	&	0	&	0	&	1	&	0	&	1	&	1	\\
\hline
\hline
\end{tabular}
\begin{tabular}{cll}
             Column       &    Notation                 &      \\
\hline
   1 &   $N$              &       Energy level reference number (row)    \\
   2 & $\tilde{E}$        &       Term value (in \cm)
\\
   3 &  $g_{\rm tot}$     &       Total degeneracy   \\
   4 &  $J$               &       Rotational quantum number    \\
   5 &  $\Gamma_{\rm tot}$&       Total symmetry in \Cvvv(M)
\\
6 &  $K$               &       Rotational quantum number, projection of $J$
onto the $z$-axis                \\
7 &  $\Gamma_{\rm rot}$& Symmetry of the rotational contribution in \Cvvv(M) \\
8 &  $L$               &       The projection of the total vibrational angular momentum                             \\

   9,10,11,12 &  $n_1-n_4$  &       Normal mode vibrational quantum numbers   \\
   13,14 &  $ L_3, L_4$ &   projections of the angular momenta corresponding to $n_3$ and $n_4$ \\
  15 &  $\Gamma_{\rm vib}$&       Symmetry of the vibrational contribution in
\Cvvv(M) \\
16 & $|C_i|^2$ & Largest contribution used in the assignment\\
 17,18,19,20,21,22 &  $s_1,s_2,s_3,b_1,b_2,b_3$  &       Local mode vibrational quantum numbers   \\
\hline
\end{tabular}
\end{table}

\begin{table}
\caption{\label{t:Transit-file} Extract from the SAlTY Transition file.}
\begin{center}

\renewcommand{\arraystretch}{1.0}
\begin{tabular}{rrc}
    \hline
    \hline
         $F$  &  $I$  & A$_{\rm IF}$ / s$^{-1}$ \\
\hline
     4220641	 &       4736989	&9.0696e-04  \\
     8442759	 &       8640461	&5.3636e-05  \\
     1269889	 &       1056999	&5.5676e-04  \\
     4631869	 &       4737012	&2.3014e-04  \\
     4632512	 &       4737066	&9.6883e-04  \\
      614599	 &        820125	&1.0712e-03  \\
     3549641	 &       3825894	&9.3653e-04  \\
     8085571	 &       7937418	&5.3630e-07  \\
     2304706	 &       2606502	&2.4236e-03  \\
     3829402	 &       3545923	&1.0250e-04  \\
     1750096	 &       1497115	&2.3840e-04  \\
      823228	 &        612463	&4.8085e-07  \\
     7589341	 &       7582878	&8.6990e-04  \\
      507260	 &        612492	&5.5009e-04  \\
     6611560	 &       6605474	&2.2927e-04  \\
     2306005	 &       2300595	&1.8031e-04  \\
     5869339	 &       5986016	&1.5085e-04  \\
      870236	 &       1057299	&4.1518e-04  \\
    \hline
    \hline
\end{tabular}

\noindent
 $F$: Upper state counting number;
$I$:      Lower state counting number; $A_{IF}$:  Einstein $A$ coefficient in s$^{-1}$.

\end{center}

\end{table}

\section{Background to the calculation}
\label{s:qn}

For this calculations we used a slightly refined version of the potential energy surface (PES)
used for the SYT line list by \citep{jt556}. This refinement was necessary because in our
semi-empirical approach the refined PESs are `effective' surfaces, inextricably connected
to the size of the basis set. To achieve the accuracy and degree of completeness required
of the SAlTY line list, the present calculations used an increased basis set, so a
refinement of the surface was necessary. It was done by performing a least square fit to
available experimental ro-vibrational energies of PH$_3$ with the rotational quantum
numbers $J=0,1,2,3,4$, mostly taken from the HITRAN database.


The PES parameters used here are given as Supplementary Material to this paper in the form
of a Fortran 95 program. It should be noted that this is also an `effective' PES and
guarantees to give accurate results only in conjunction with the same method and basis
set used to produce it with.

As with the SYT PH$_3$ line list, the variational rotation-vibration program suite TROVE
\citep{07YuThJe.method} was employed for all nuclear motion calculations for SAlTY. The
application of variational methods to polyatomic molecules requires increasingly large
Hamiltonian matrices to be diagonalised, which is very computationally demanding and
until recently prohibitively so. With the power and parallelism of modern computers, it is now
possible to use this method for the production of accurate spectra.

To accommodate the higher demands of the present line list, a larger basis set was used,
with a corresponding higher polyad number. The polyad number controls the size of the
basis sets at all contraction steps using the polyad-truncation scheme
\citep{08OvThYu2.PH3}, defined by
\begin{equation}\label{e:polyad-P}
    P =  2 (s_1 + s_2 +s_3) + b_{1} + b_{2} + b_{3} \le P_{\rm max},
\end{equation}
where $s_i$ and $b_i$ are the primitive quantum numbers associated
with the basis functions, $\phi_{s_i}$ and $\phi_{b_i}$, for the
stretching modes and the bending modes, respectively. Here TROVE
\citep{07YuThJe.method} constructs a synthetic line list for
$^{31}$PH$_3$ in its ground electronic state. To this end the \schr\
equation for the rotation-vibration motion of nuclei is solved to
obtain eigenvalues (ro-vibrational energies) and eigenfunctions
(nuclear motion wavefunctions). The latter are necessary for
ro-vibrational averaging of the dipole moment of the rotating molecule
and thus to compute the transitional probabilities, usually expressed
in terms of the Einstein coefficients or line strengths following the
description by \cite{05YuThCa.method}. Our basis set contains two
contributions, (i) all basis functions with the primitive quantum
numbers satisfying $P \le 16$ and (ii) stretching functions ranging up
to $P=20$ but with some high $P$-polyad ($P\ge 17$) stretching
contributions that couple all three stretching modes removed.
Increasing the polyad number is computationally costly, and as such
the increased coverage of the stretching excitations only was
motivated by assumption that the stretching excitations produce the
strongest transitions.  The larger basis set guarantees a better
convergence for the present calculations, but requires the PES to be
refined to the new basis set, as discussed above. The $J=0$
Hamiltonian matrices were then constructed, and the empirical basis set
correction scheme (EBSC) of \citet{jt466} was applied. Here, band
centre values from the ro-vibrational calculations are replaced with
extremely accurate corresponding experimental values, or deviated
towards these values. This is performed iteratively, until the entire
band is considered to be optimal.  At the end of this correction, the
rms deviation for the bands whose centre had been replaced went from
an rms deviation of 0.02 cm$^{-1}$ in SYT to 0.012 cm$^{-1}$ in SAlTY.
 This improvement may reflect 
the better use of EBSC, but overwhelmingly is due to the use of
the enhanced PES.  Table \ref{BandCentres} gives a band-by-band 
summary of this
improvement. The 2$\nu_4$ and $\nu_2 + \nu_4$ bands were particularly responsive to this
process. Further details of the TROVE
computational procedure are given by \cite{jt556}.

Table \ref{BandCentres} also compares the integrated band intensity
computed for the SYT and SAlTY line lists with those obtained from HITRAN 2012. 
In each case, the band intensity was computed by
explicit summation of the intensities at 296 K of all lines
within a band. The two $2\nu_4$ bands were considered together since it is
difficult to disentangle their transitions. The band intensities for
the two computed lists are very similar. They are both in reasonable agreement
with, but somewhat larger than, HITRAN; 
given the greater completeness of the computed linelists, it is
to be expected that this method would yield somewhat larger band intensities.

For the rotational spectra from CDMS the integrated intensity is
1.214$\times10^{-18}$ cm/molecule which is very similar to SAlTY's
1.218$\times10^{-18}$.  One further test on the intensities was
performed. Given that the intensity of weak lines can be highly
sensitive to the choice of wavefunction \citep{jt573}, and hence the
underlying PES, an explicit comparison was made between SYT and SAlTY of
the intensities of the weak, forbidden lines with $\Delta K =3$. These were found to be very little changed for the individual transitions
inspected, with the integrated intensity
for all these transitions going from $5.53\times10^{-19}$ cm / molecule
in SYT to $ 5.60\times10^{-19}$ cm / molecule in SAlTY.  In practice
variational procedures, such as those use here, have long been used to
get reliable predicted intensities for such transitions \citep{jt89,jt428}.

\begin{table}
\small \tabcolsep=5pt
\renewcommand{\arraystretch}{1.1}

\caption{Observed bands centres, from HITRAN \citep{jt557}, and standard deviation, $\sigma$,
  with which the TROVE calculations reproduce the terms within each band.
The first three columns give integrated band intensity for each band calculated for SYT, SAlTY and HITRAN.}

\begin{tabular}{lrcccccc}
\hline\hline
\multicolumn{1}{l}{Band} &\multicolumn{1}{c}{Band Centre} & \multicolumn{2}{c}{$\sigma$ (cm$^{-1}$)} & & \multicolumn{3}{c}{Band Intensity (cm / molecule)}  \\ 
\cline{3-4} \cline{6-8}
 &\multicolumn{1}{c}{ (cm$^{-1}$)} & \multicolumn{1}{c}{SYT} & \multicolumn{1}{c}{SAlTY} &  & \multicolumn{1}{c}{HITRAN} & \multicolumn{1}{c}{SYT} & \multicolumn{1}{c}{SAlTY}\\
\hline
$\nu_{2}$	&$992.135$ &	0.020	&	0.015	&&	 $3.087\times10^{-18}$ &                  $3.264\times10^{-18}$ & $3.261\times10^{-18}$\\
$\nu_{4}^1$&$1118.307$ &	0.005	&	0.003	&&	 $3.149\times10^{-18}$&          $3.436\times10^{-18}$ & $3.436\times10^{-18}$\\
$2\nu_{2}$	&$1972.571$&	0.007	&	0.004	&&	 $9.390\times10^{-21}$&          $1.414\times10^{-20}$ & $1.468\times10^{-20}$\\
$\nu_{2}+\nu_{4}^1$	&$2108.152$ &	0.034	&	0.013	&& $9.504\times10^{-20}$&   $1.484\times10^{-19}$&  $1.487\times10^{-19}$\\
$2\nu_{4}^0$	&$2226.835$ &	0.010	&	0.008	&&	 $4.689\times10^{-19}$&            $7.684\times10^{-19}$& $7.590\times10^{-19}$\\
$2\nu_{4}^2$	&$2234.915$ &	0.014	&	0.005	&&            combined above & combined above& combined above\\
$\nu_{1}$	&$2321.121$&	0.012	&	0.008	&&	$4.926\times10^{-18}$&                    $6.091\times10^{-18}$ & $6.127\times10^{-18}$\\
$\nu_{3}^1$	&$2326.867$ &	0.013	&	0.008	&&	 $1.454\times10^{-17}$&            $1.650\times10^{-17}$ & $1.648\times10^{-17}$\\
$3\nu_{2}$	&$2940.767$&	0.046	&	0.032	&&	 $ 6.725\times10^{-21}$ &            $ 2.127\times10^{-20}$ & $ 2.129\times10^{-20}$\\
$\nu_{2}+2\nu_{4}^0$	&$3214.936$ &	0.024	&	0.017 &&  $3.612\times10^{-21}$ &       $7.478\times10^{-21}$ & $6.996\times10^{-21}$\\
Overall & & 0.020	&	0.012	&	\\

\hline\hline

\label{BandCentres}
\end{tabular}
\end{table}


As well as employing a larger basis set than before, this line list extends SYT by using
(i) a larger energy level range of E$_{\rm max}\leq$ 18 000 cm$^{-1}$(instead of $\leq$
12 000 cm$^{-1}$), (ii) wider frequency range of 0 -- 10 000~cm$^{-1}$ (instead of  0 --
8 000~cm$^{-1}$) and (iii) rotational excitations considered up to $J_{\rm max}=46$
(instead of 34). This expansion is necessary to guarantee accuracy and completeness at
high temperatures, but it makes the calculation of the line list much more
computationally demanding.

The diagonalisation of the matrices involved in the variational procedure corresponding
to high values of $J$ is the most computationally demanding part of a line list
calculation, as it requires substantial memory, long runs and requires MPI to diagonalise
efficiently. A variety of strategies were used to deal with this. The dimension of the
matrices of the symmetries scales approximately as 1:1:2, for $A_{1}$, $A_{2}$ and $E$,
respectively.

\begin{figure}
\centering
{\leavevmode \epsfxsize=14.0cm \epsfbox{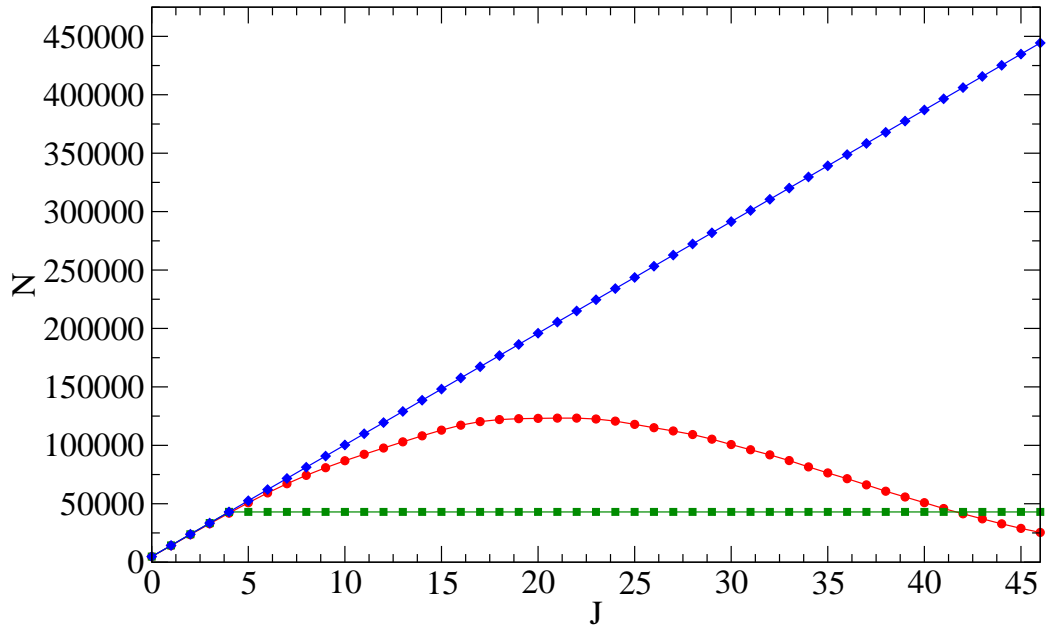}
\label{matrix}}
\caption{Dimensions of the $E$-symmetry matrices with $J$ (blue diamonds),
the corresponding number of eigenvalues below 18~000~cm$^{-1}$ (red circles) and number of non-zero elements on each row (green squares).  }

\end{figure}

The largest matrix to be diagonalised for SAlTY ($J=46$, $E$ symmetry) has dimensions of
444~726. As can be seen from Fig.~\ref{matrix},  the size of the ro-vibrational matrices
scales linearily with $J$, but the number of non-zero elements and the number of
eigenvalues under the energy threshold for the line list (8 000 cm$^{-1}$) do not. The
size of the matrices grows roughly with \( N_{J=0}^{\Gamma} \times (2J + 1)\), where \(
N_{J=0}^{\Gamma}\) is the dimension of the matrix block for  \(J=0\). For the symmetry
with the biggest matrices, $\Gamma = E$, \( N_{J=0}^{E}\) = 4778.

The initial (low $J$) matrices were sufficiently small to employ the standard LAPACK
eigensolver DSYEV  \citep{99AnBaBi.method} to solve the full eigenvalue problem, and this
was used for all $J \leq 16$ and 9 ($A$ and $E$ symmetries, respectively). As the
matrices increased in size, memory constraints required the MPI version PDSYEVD of this
eigensolver to be used on large shared memory systems (COSMOS) in order to cope with the
size of these matrices. We also successfully explored the openmp PLASMA library
\citet{13KuLuYa.method} for the matrices with dimensions between 100~000 and 200~000.

 Using these three eigensolvers, all eigenvalues and eigenvectors were computed
for $J \leq 46$, using the energy threshold of 18 000~cm$^{-1}$, totalling 7~480~690
energy levels.


A high quality dipole moment surface (DMS) is a prerequisite for correctly computing the
transition moments used to generate the Einstein $A$ coefficients and in turn accurate
line intensities; \citet{jt573} gives a discussion of the issues involved in this. The
DMS used here is completely theoretical, an approach that has been shown to yield results
competitive with experiment \citep{jt156}, and has already been used in the SYT line list
where it provided intensities in good agreement with experiment. It is a six-dimensional
{\it{ab initio}} (CCSD(T)/aug-cc-pVTZ) electric dipole moment surface, calculated on a
grid of 10~080 molecular geometries, and it is described in further detail elsewhere
\citep{06YuCaTh.PH3}. To reduce the size of the computation, only eigen-coefficients of
the upper states with a magnitude larger than 10$^{-14}$ were selected.

\begin{figure}

\centering
{\leavevmode \epsfxsize=14.0cm \epsfbox{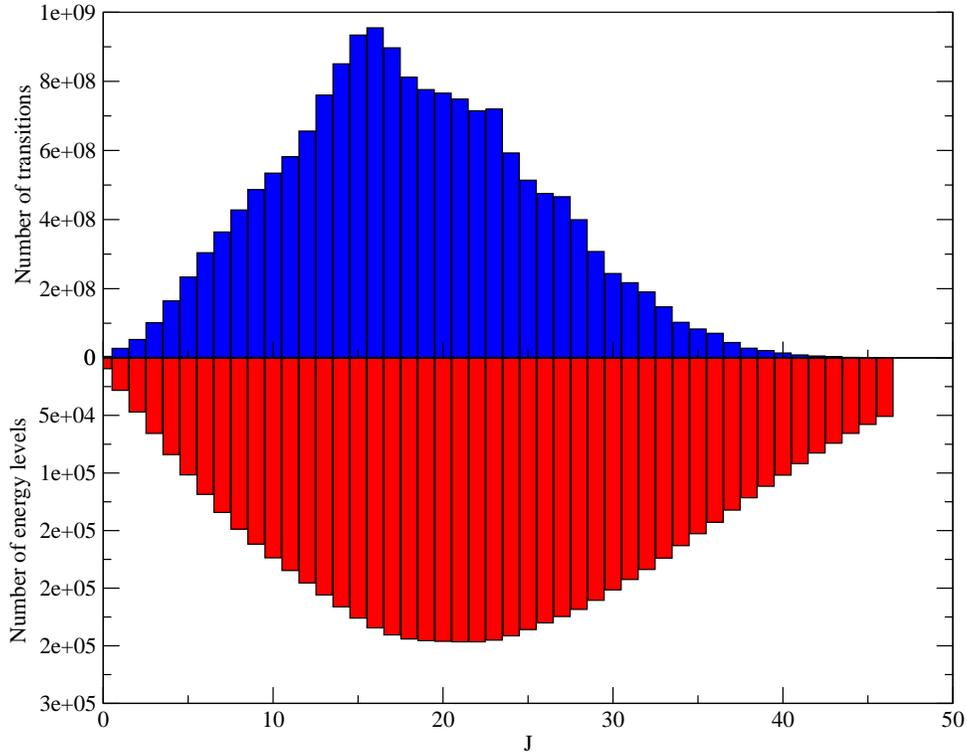}
\label{f:lines:J}}
\caption{Number of energy levels in each rotational quantum number, $J$, and transitions between $J$ and $J$,$J+1$ in the SAlTY line list, summed over all symmetries.}
\end{figure}

Despite matrix diagonalization utilising the most computational
resources, it is the computation of transition intensities that
actually dominates the total computer time. However this step has the
advantage that it is easily paralellised and spread over many
processors, and indeed computer systems. As shown in
Fig.~\ref{f:lines:J}, the number of transitions between energy levels
peaks between $J=15$ and $18$, so this was the region that
corresponded to the most expensive part of the intensity calculations,
accounting for about 17\% of the total transitions. The number of
transitions does not correlate smoothly with $J$ pairs due to the
different way $A$ and $E$ symmetries are affected by $J$, particularly
for values of $K$ that are multiples of 3.

Employing CPUs available to us at the HPC centres DiRAC@Darwin and DiRAC@COSMOS, TROVE
was used to calculate the intensities across almost all symmetries and $J$ values.
However this proved slow and computationally expensive, so for some of the particularly
computationally demanding sections, a newly-developed graphical processor unit (GPU)
implementation of the TROVE program was employed. This allowed us to use multiple GPUs to
compute multiple initial states simultaneously, effectively reducing computing time by
over an order of magnitude. This was used to compute the absolute intensities for most
$A$ symmetry transitions with $J\geq20$, on the Emerald cluster. The algorithm employed
will be described elsewhere \citep{15AlYuTe.PH3}.

After this work was completed, \citet{14NiReTy.PH3} published a detailed analysis of 
{\it ab initio} methods for calculating the phosphine DMS. In general the agreement
between their predictions and ours is very good.

\section{Line list validation and temperature dependence}
\label{s:analysis}

Line lists do not specify a temperature, since the Einstein coefficients for the
transitions are independent of temperature. SAlTY can only be considered a `hot' line
list because its energy and rotational excitation thresholds ensure that all those states
that are significantly populated up to $T=T_{\rm max}$ are used to produce the catalogue
of transitions.

To model the spectrum of phosphine at different temperatures, the line intensities are
calculated using:
\begin{equation}
I(f \leftarrow i) = \frac{A_{if}}{8\pi c}g_{\rm ns}(2
J_f+1)\frac{\exp\left(-\frac{E_{i}}{kT}\right)}{Q\;
\tilde{\nu}_{if}^{2}}\left[1-\exp\left(\frac{-hc\tilde{\nu}_{if}}{kT}
\right)\right],
\label{eq:intens}
\end{equation}
where \(k\) is the Boltzmann constant, \(T\) the absolute temperature and \(g_{\rm ns}\)
is the nuclear spin statistical weight factor. \(Q\), the partition function, is given
by:
\begin{equation}
  Q = \sum_{i} g_{i}\exp\left({\frac{-E_{i}}{k T}}\right),
  \label{eq:part}
\end{equation}
where \(g_{i}\) is the degeneracy of a particular state \(i\) with
energy \(E_{i}\). For PH$_{3}$, \(g_{i}\) is \(g_{\rm ns}(2 J_i + 1)\) with
$g_{\rm ns} = 8$ for all \(A_{1}\), \(A_{2}\)  and \(E\) symmetries.

Previous work has looked in depth at the temperature dependent partition function of
phosphine \citep{jt571}. This work produced accurate and fully converged values for the
partition function for temperatures below 3000~K by the explicit summation of theoretical
rotation-vibration energy levels of the molecule. The convergence dependence on
temperature arises due to the growing contribution higher states make towards the value
of the partition function as temperature increases. To achieve this convergence, 5.6
million energy levels with high accuracy were used, combined with an additional 145
million calculated with an accurate vibrational component but with the rotational
contribution estimated by a rigid rotor approximation, and consequently with a much
decreased level of accuracy. However, the rotational states are anchored to the
vibrational states, so an accurate vibrational description means that the energy level
clusters remain valid \citep{05YuThPa.PH3}, even if the degradation of accuracy from the
rotational component makes the energies within the cluster only approximate.
Consequently, their collective contribution to the partition function and related
properties remains correct.

\begin{figure}
\centering
{\leavevmode \epsfxsize=12.0cm \epsfbox{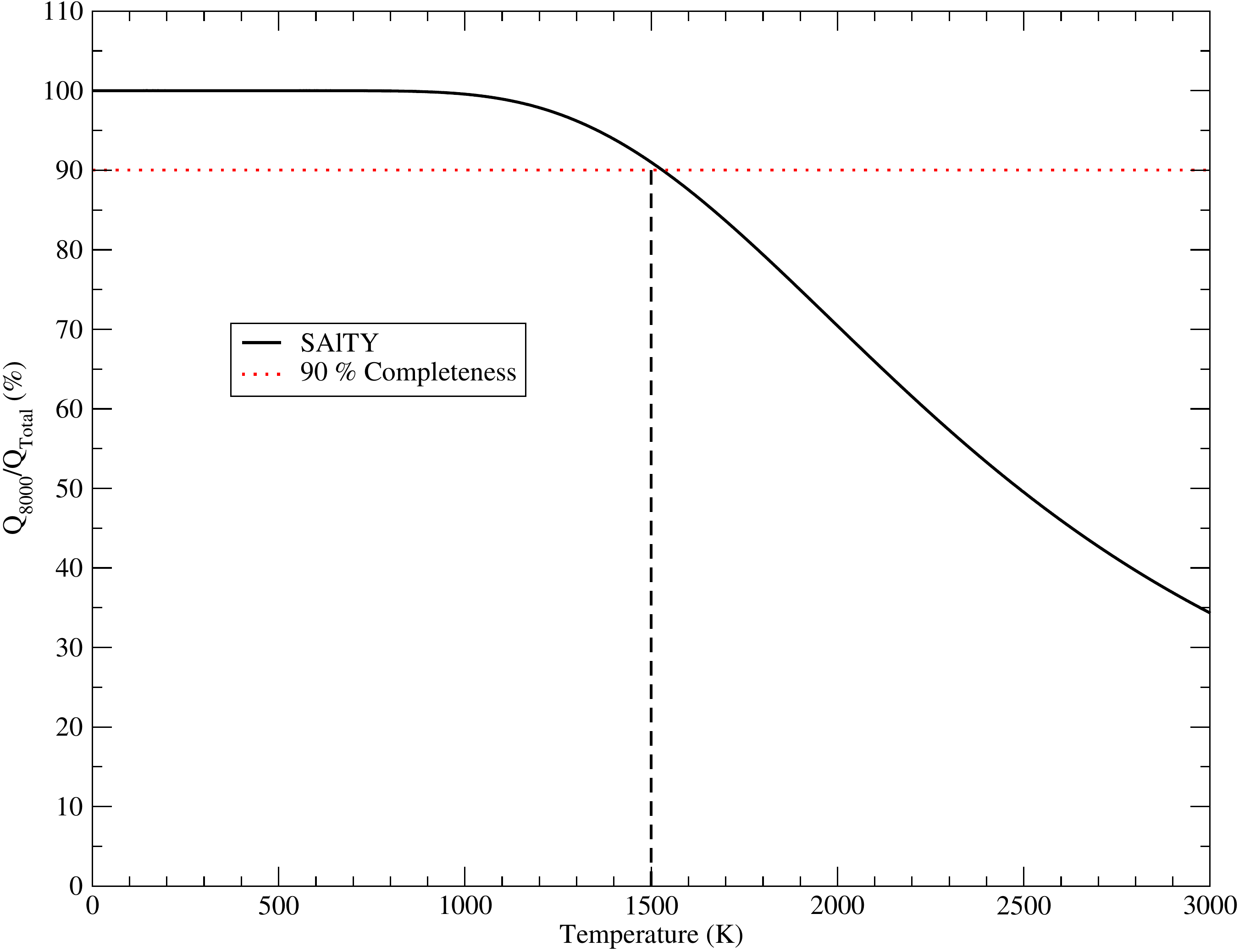}
\label{Qratio}}
\caption{Ratio of effective partition function used in SAlTY, $Q_{8000}$, to the converged value, $Q_{\rm total}$, calculated in \citet{jt571}.
This ratio gives a measure of completeness of SAlTY as a function of temperature.}

\end{figure}

The $T_{\rm max}$ for which molecular data is complete to a satisfactory degree can be
found by computing the temperature-dependent partition function using only the energy
levels under the lower energy threshold considered in the line list (here 8 000~cm$^{-1}$), and then
comparing its value to that of the complete partition function. By considering the ratio
$Q_{\rm limit}/Q_{\rm total}$, where $Q_{\rm total}$ is the converged partition function
value calculated by explicitly summing over all energy levels and $Q_{\rm limit}$ is the
partition function calculated using only those levels with energies up to SAlTY's
threshold of 8 000~cm$^{-1}$. Fig.~\ref{Qratio} shows SAlTY's completeness with
increasing temperature. As can be seen from the Table \ref{Completeness}, SAlTY is over
90\% complete for temperatures below 1500~K, but quickly becomes depleted at higher
temperatures. It is possible to use SAlTY to model temperatures over 1500~K by using the
percentual loss of completeness to estimate the proportion of opacity missing from a
spectrum \citep{jt181}, but it is recommended that 1500~K is taken as a soft limit to
the applicability of the SAlTY line list.

\begin{table}
\caption{\label{Completeness} Maximum temperatures for which the SAlTY line list is percentually complete.}
\begin{center}
\small \tabcolsep=14pt
\renewcommand{\arraystretch}{1.1}
\begin{tabular}{rc}
    \hline
    \hline
         $T_{\rm max}$ (K)  &   Completeness $\%$  \\
\hline
           1014 &     100  \\
            1146 &     99  \\
             1797 &     80  \\
              1500 &     91  \\
         2000 &     70  \\
             2500 &     50  \\

    \hline
    \hline
\end{tabular}

\end{center}

\end{table}

\begin{figure}

\centering
{\leavevmode \epsfxsize=11.0cm \epsfbox{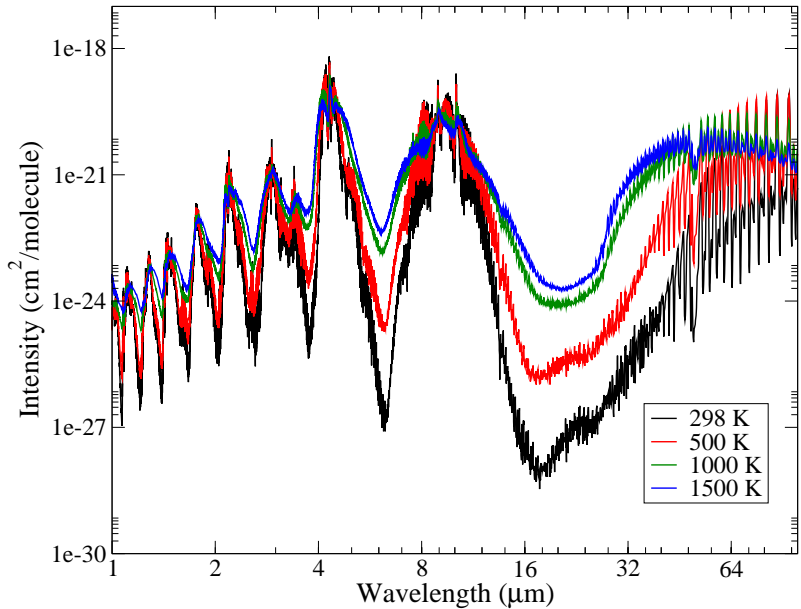}
\caption{Overview of the full SAlTY spectrum of PH$_{3}$ for $T$ = 300, 500, 1000 and 1500~K, absorption cross-sections (cm$^2$/molecule) with HWHM = 0.5 cm$^{-1}$. Looking at the minimum of the spectra, the cross-sections are ordered in increasing temperature. }
\label{PH3_overview}}
\end{figure}
\begin{figure}
\centering
{\leavevmode \epsfxsize=11.0cm \epsfbox{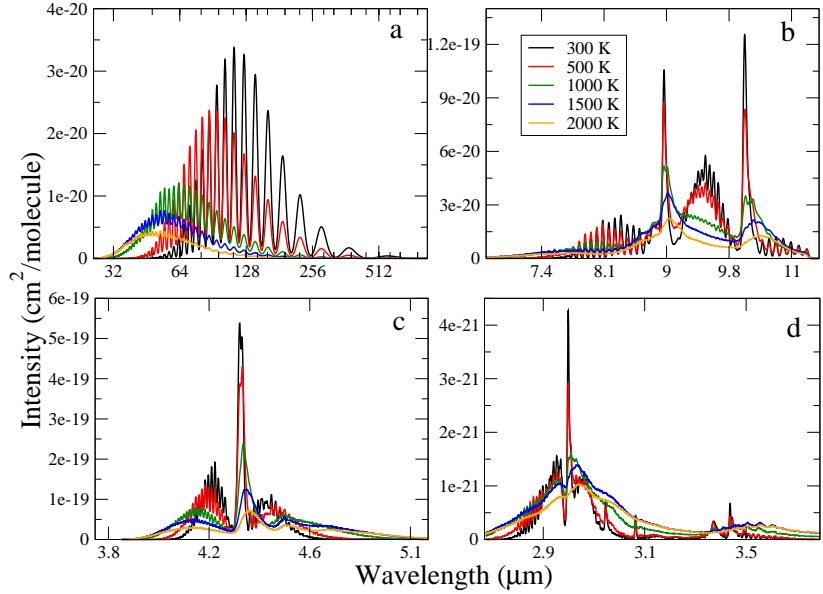}

\caption{SAlTY absorption spectra of PH${_3}$ for $T$ = 300, 500, 1000, 1500 and 2000~K, convoluted with a Gaussian profile, HWHM = 2 cm$^{-1}$, for the 30({\bf{a}}), 10({\bf{b}}), 5({\bf{c}}) and 3({\bf{d}}) $\mu$m regions}
\label{PH3_zooms}}

\end{figure}

Using the equations above and the present line list, multiple
cross-sections were simulated for PH$_3$. Figures ~\ref{PH3_overview}
and ~\ref{PH3_zooms} illustrate the temperature dependence of
phosphine for $T$ = 300, 500, 1000, 1500 and 2000 K. It can be seen
that the spectra at higher temperatures have less features and a loss
of sharp $Q$-branches.  Figure \ref{PH3_zooms}, part {\bf{c}}, covers
the wavelength region 3.8 -- 5.1 $\mu$m which, as mention above, is of
particular importance for the spectral characterisation of brown
dwarfs and gas giants.

\begin{figure}

\centering
{\leavevmode \epsfxsize=13.0cm \epsfbox{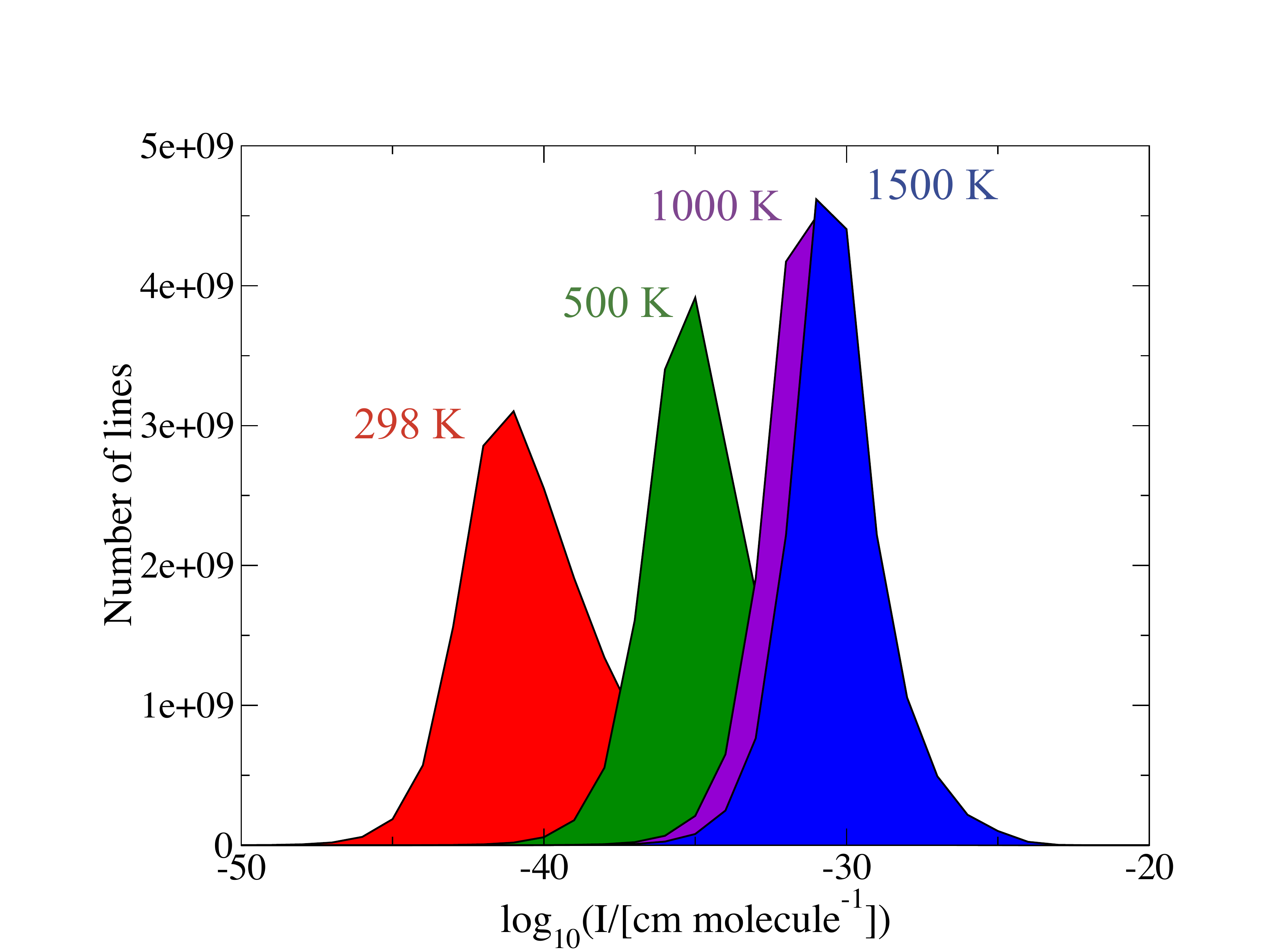}

\caption{Number of intense lines as a function of intensity
for different temperatures.
The $x$-axis gives the log of the intensity in cm/molecule,
while the $y$-axis represents the number of transitions per each 10$^{x}$
cm/molecule bin.}
\label{f:nlines:temp}}
\end{figure}

Fig.~\ref{f:nlines:temp} demonstrates how the density of lines per an absorption
intensity $I_{if} = A \times 10^x$ unit changes with temperature, covering the whole
wavenumber range 0--10~000 cm$^{-1}$. As the temperature rises, the number of intense
lines increases but the range of intensities in the spectrum becomes narrower. The
Gaussian-like intensity distributions peak at $I$ =$10^{-41}$, $=10^{-35}$, $=10^{-31}$,
and $=10^{-31}$ for $T$ = 298, 500, 1000, and 1500 K, respectively. This is different
from the intensity distribution found for CH\4\ by \citet{jt564}, where the proportion of
strong lines was found to be much larger.

\subsection{Comparison with experiment}
\label{s:HITRAN}

\begin{figure}

\centering {\leavevmode \epsfxsize=14.0cm \epsfbox{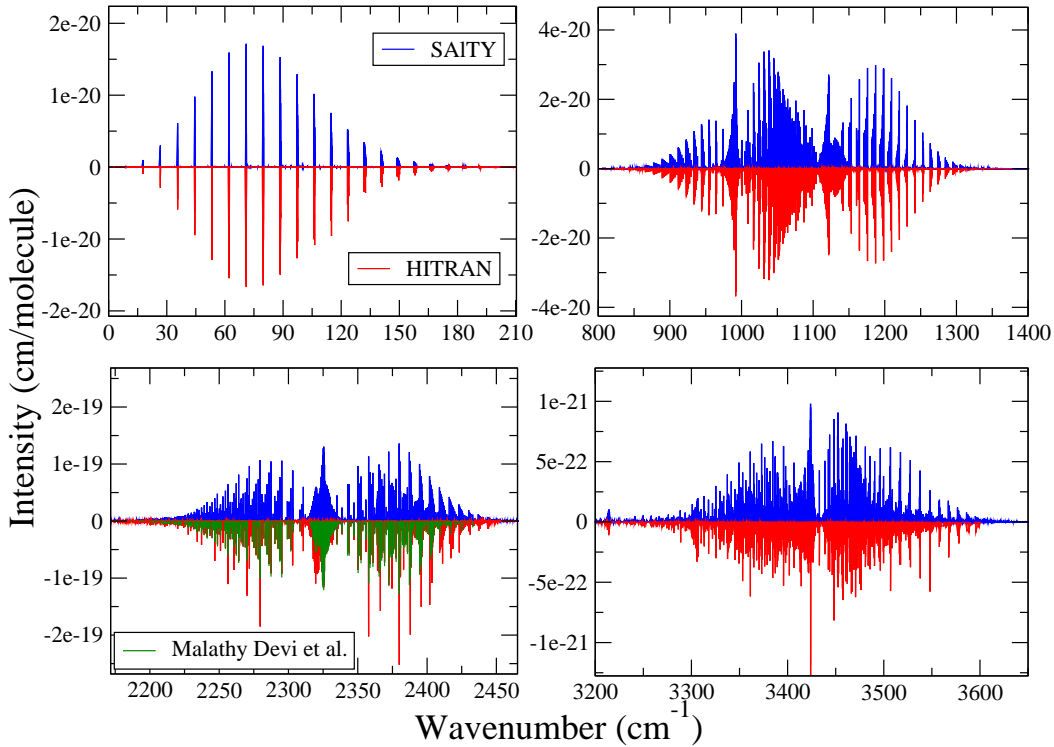}
 \caption{Comparison of the SAlTY line list with the phosphine
data from HITRAN \citep{jt557} at room temperature; also
shown are the recent results of \citet{14DeKlSa.PH3}. \label{f:HITRAN}}
\label{hitran_comparison}}
\end{figure}

\cite{jt556} offers a detailed comparison of SYT to the existing
experimental data, where it was found that it replicates very well the
observed phosphine spectra at room temperature, with a maximum rms
deviation from CDMS \citep{05MuScSt.db, 13Muxxxx.PH3} of 0.076
cm$^{-1}$ for the rotational spectrum and of 0.23 cm$^{-1}$ from
HITRAN \citep{jt557}. As can be seen from
Fig.~\ref{hitran_comparison}, the SAlTY line list is also in excellent
agreement with HITRAN (and incorporated CDMS data), and it is expected
that SAlTY will have slightly lower rms deviations from experiment at
this temperature.

\begin{figure}

\centering {\leavevmode \epsfxsize=10.0cm \epsfbox{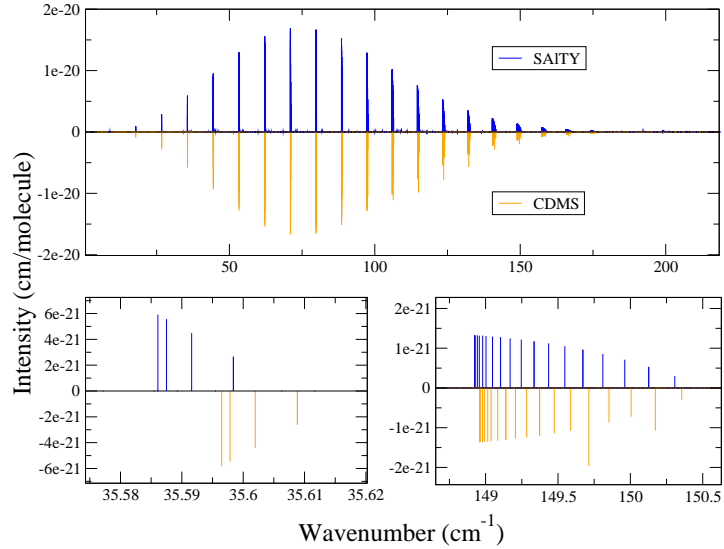}
 \caption{Comparison of the SAlTY line list with the most recent phosphine
data from CDMS at room temperature \citep{13Muxxxx.PH3}. }
\label{cdms}}
\end{figure}

The most recent CDMS update \citep{13Muxxxx.PH3} contains a very comprehensive description of the pure rotational band of phosphine. Figure \ref{cdms} demonstrates that SAlTY is in excellent agreement with the CDMS data. The transitions in the bottom left box shows a deviation of approximately 0.01 cm$^{-1}$, while the right most box, at higher wavenumbers, shows a deviation of approximately 0.04 cm$^{-1}$.

The apparent intensity disagreements with HITRAN occur because of an
overestimation of the intensity of some of the degenerate $A_{1}
\leftrightarrow A_{2}$ transitions which are in such close proximity
that they have been perceived as one doubly strong transition in the
experimental data. This has been partially resolved by
\citet{14DeKlSa.PH3}, who corrected these intensities for the region
1950 to 2450 cm$^{-1}$. The bottom left box in figure
\ref{hitran_comparison} includes this data, and it can be seen SAlTY
is in much better agreement with it than with the HITRAN data. We
recommend that the \citet{14DeKlSa.PH3} data should be included in  the
next release of HITRAN.

\begin{figure}

\centering {\leavevmode \epsfxsize=16.0cm \epsfbox{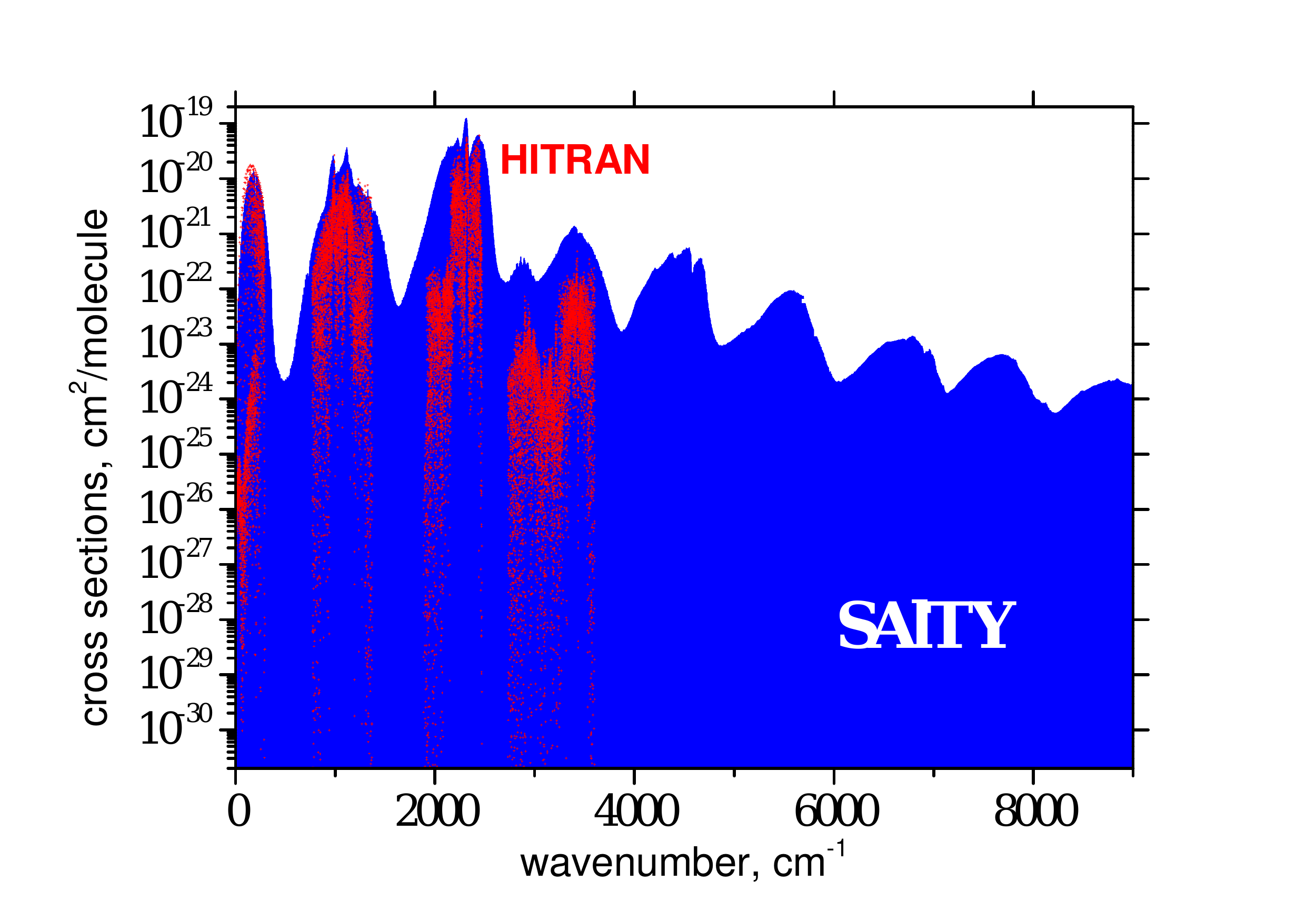}
 \caption{Comparison of the SAlTY line list with the phosphine
data from HITRAN \citep{jt557} at 1500 K.}
\label{hightemp_overview}}
\end{figure}

The data in HITRAN are generally very accurate for room temperature simulations. 
However, if the current HITRAN database is used to simulate high temperature spectra,
the differences become much more striking, as can be seen in Fig.~\ref{hightemp_overview}.

\begin{figure}

\centering {\leavevmode \epsfxsize=12.0cm \epsfbox{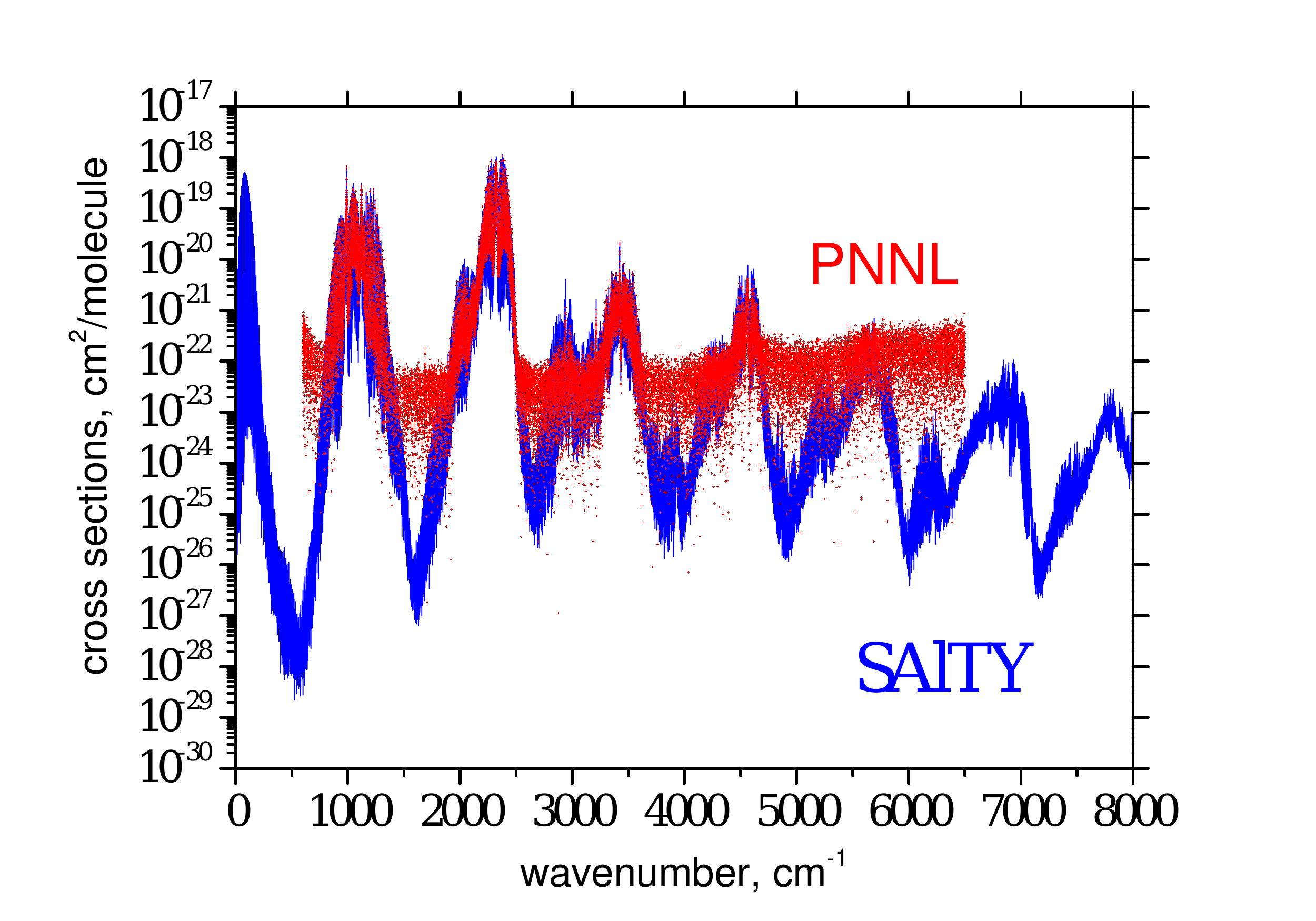} 
\caption{Comparison of the SAlTY absorption cross-sections for PH\3\  with PNNL and
HITRAN at $T=50$~C, HWHM = 0.076 cm$^{-1}$.   }
\label{f:PNNL}}
\end{figure}

The only available data above 300 K comes from the Pacific Northwest
National Laboratory (PNNL) who provide cross-sections for PH$_{3}$ up
to temperatures of 323 K \citep{PNNL}. Comparisons with SAlTY for $T=50$~C, or 323 K, can be seen in Fig.~\ref{f:PNNL}. The
PNNL data is approximately $8\%$ weaker than SAlTY's, but both are otherwise in good agreement. Given its larger density and coverage, it is recommended that
SAlTY is used to simulate cross-sections, even at low temperatures.

\section{Conclusion}
\label{s:conclusion}

An accurate and comprehensive line list for phosphine is presented, which has been called
SAlTY. It contains 16.8 billion transitions between 7.5 million energy levels and it is
suitable for simulating spectra up to temperatures of 1500~K.

Its only limitations are: upper states with energies above 18\,000~cm$^{-1}$ are
excluded, and there is an effective short-end wavelength cut-off of 1~$\mu$m,
which is an unimportant
region for PH$_{3}$. SAlTY improves on the previous room temperature phosphine line list
SYT in terms of the size of the basis set, corresponding refined potential energy surface
and increased spectral range. It is therefore recommended that SAlTY is used for all
applications, even at low temperatures where SYT would provide reasonable results.

The tunnelling effect present in ammonia is predicted in phosphine but has yet to be
observed in phosphine \citep{81BeBuPo.PH3}, due to its much higher inversion barrier
(12~300 cm$^{-1}$). This tunnelling effect is not considered here, but rough preliminary
calculations have predicted the position, size and intensity of transitions from split
energy levels. Further work on this effect will be published elsewhere.

The SAlTY line list is presented as the latest in a series of molecules as part of the
ExoMol project, which aims to provide comprehensive line lists for every molecule
relevant to the characterisation of the atmospheres of cool stars and exoplanets. SAlTY
is freely available online in full or filtered by wavelength and intensity, from the
ExoMol website. The line list in its entirety is very large but can be made more
manageable by using cross sections \citep{jt542} (available at ExoMol) or
$k$-coefficients \citep{96IrCaTa.method}.

\section*{Acknowledgements}

This work was supported by the ERC Advanced Investigator Project 267219. The research
made use of the DiRAC@Darwin, DiRAC@COSMOS and the EMERALD HPC clusters. DiRAC is the UK
HPC facility for particle physics, astrophysics and cosmology and is supported by STFC
and BIS. The COSMOS Shared Memory system at DAMTP, University of Cambridge operated on
behalf of the STFC DiRAC HPC Facility. This equipment is funded by BIS National
E-infrastructure capital grant ST/J005673/1 and STFC grants ST/H008586/1, ST/K00333X/1.
The EMERALD High Performance Computing facility is provided via the Centre for Innovation
(CfI). The CfI is formed from the Universities of Bristol, Oxford, Southampton and UCL in
partnership with STFC Rutherford Appleton Laboratory. We thank James Briggs (COSMOS) and
Cheng Liao (SGI) for their help with the PLASMA eigensolver.

The authors would also like to thank Duncan A. Little, Hannah Lees, Sally Hewett, Helen Parks, Antonio Silva,
Antonio S. Silva and Peter J. Day for their support and contribution.

\label{lastpage}

\bibliographystyle{mn2e}


\end{document}